\documentclass[prb,twocolumn,amsmath,amssymb,eqsecnum,preprintnumbers]{revtex4}
\usepackage[T1]{fontenc}
\usepackage[latin9]{inputenc}
\usepackage{color}
\usepackage{graphicx}
\usepackage{amsmath}
\usepackage{amssymb}
\usepackage{esint}

\makeatletter
\@ifundefined{textcolor}{}
{%
 \definecolor{BLACK}{gray}{0}
 \definecolor{WHITE}{gray}{1}
 \definecolor{RED}{rgb}{1,0,0}
 \definecolor{GREEN}{rgb}{0,1,0}
 \definecolor{BLUE}{rgb}{0,0,1}
 \definecolor{CYAN}{cmyk}{1,0,0,0}
 \definecolor{MAGENTA}{cmyk}{0,1,0,0}
 \definecolor{YELLOW}{cmyk}{0,0,1,0}
}

\def\be{\begin{equation}}
\def\ee{\end{equation}}
\def\bea{\begin{eqnarray}}
\def\eea{\end{eqnarray}}
\def\bse{\begin{subequations}}
\def\ese{\end{subequations}}

\usepackage{dcolumn}% Align table columns on decimal point
\usepackage{bm}% bold math
\makeatother

\begin{document}

\title{Universal aspects of the
structural glass transition from density functional theory }

\author{T.R. Kirkpatrick$^{1}$ and D. Thirumalai$^{2}$}

\affiliation{$^{1}$Institute for Physical Science and Technology, University of Maryland, College Park, MD 20742\\
 $^{2}$Department of Chemistry, University of Texas at Austin, Austin,
TX 78712}

\date{\today}
\begin{abstract}

The random first order transition (RFOT) of the structural glass transition (SGT) can be formulated as a density functional theory (DFT). An important feature of the complete RFOT theory is that it has at least two distinct transition temperatures, one of which is a dynamical transition signaling loss of effective ergodicity, and the other is an equilibrium ideal glass transition. The dynamical transition exists only in mean-field-like theories and becomes a rounded transition when activated transport processes become prominent. Using scaling and renormalization group (RG) ideas, universal aspects of both of these transitions are discussed. Particular attention
is paid to the coherence, or correlation, length associated with these
transitions in the RFOT. We also derive universal scaling relations for several experimentally measurable
quantities that are valid near the ideal glass transition. 
In particular, activated scaling ideas are used to obtain a scaling equation for the non-linear
 structural glass susceptibility as the ideal glass transition is
approached. Important finite corrections to the ideal glass transition
temperature are also discussed. Our work provides additional criteria for assessing the validity of RFOT theory of the liquid to glass transition, and provides a firm theoretical foundation  for analyzing experimental data on the temperature dependence of the relaxation times.

\end{abstract}

%\pacs{64.60.Bd, 64.70.kj}

\maketitle

\section{Introduction}
 \label{sec:I}

When undercooled rapidly almost all liquids undergo a transition to
a glassy state, characterized by a dramatic increase in viscosity
over a narrow temperature range. Because of the central importance
of this phenomenon in condensed matter physics, material science,
and biology there has been considerable interest in understanding
the characteristics of the structural glass transition (SGT) \cite{Kirkpatrick95Transport,Cavagna09PhysRep,Parisi10RMP,Berthier11RMP,Biroli12, KirkpatrickRMP2015}.
In the late 1980s, we along with Wolynes developed the random first order phase transition
(RFOT) \cite{Kirkpatrick89PRA} theory of the glass transition \cite{Kirkpatrick87PRL,Kirkpatrick87PRB,Kirkpatrick87PRBPotts,Kirkpatrick88PRBPotts,Kirkpatrick95JPhysiqueI}
to explain many the unique aspects of the structural glass transition
(SGT) problem. Although initially motivated by precise solutions of
a class of infinite range spin glass models (see \cite{Kirkpatrick95Transport} for a short review), we subsequently used a density functional approach and established
that the generic aspects of RFOT can be found in systems even without quenched
disorder \cite{Kirkpatrick89JPhysA}.
It was further argued that the structural glass transition is an example
of a RFOT. The emergence of RFOT ideas in a theory without quenched disorder is an important development because it showed that liquid state theories could form the basis for describing glassy states.  Indeed, the RFOT has been much discussed in this
context and in many ways clarified and further extended \cite{Mezard96JPhysA,Mezard00JPhysCondMatt,Bouchaud04JCP,Toninelli05PRE}.

There are necessarily at
least two \cite{Kirkpatrick87PRL,Kirkpatrick87PRB}  distinct
transitions within the RFOT theory in the mean-field limit. When the temperature is decreased,
the liquid undergoes the so-called dynamical transition at a temperature,
$T_{d}$. This transition is, in general, not a true equilibrium phase
transition, although it is related to the topology of the
state space of an equilibrium system in that at $T_{d}$ there is
a breakdown in effective ergodicity. Technically, it can be described
by using either an equilibrium approach \cite{Kirkpatrick89JPhysA,Franz12PNAS,Franz13JCP}, or a dynamical approach \cite{Leutheusser84PRA,Bengtzelius84JPhysC,Kirkpatrick88PRBPotts,Thirumalai88PRB}.
The dynamical approach is consistent with the mode coupling
theory glass transition \cite{Leutheusser84PRA,Bengtzelius84JPhysC,Das04RMP} in that both the theories show that the nature of relaxation dynamics changes at $T_d$. The dynamical order parameter of the theory 
is a freezing of the density fluctuations %\footnote{Approximate solutions to certain non-linear
%oscillator equations lead to glassy like behavior that looks like
%the MC dynamical transition. An exact solution of the starting equation
%shows that this behavior is an artifact of the approximations. This
%is to be contrasted with various exactly soluble models which do show
%the dynamical transition discussed here, and to various systematic
%liquid state theories which also show such a transition.}. 
To further characterize
the nature of the transition at $T_{d}$, we introduced \cite{Kirkpatrick88PRA}
a non-linear susceptivity or four point correlation function ($\chi_{NL}(t)$ where $t$ is time),
which is the natural susceptibility for this transition \cite{Kirkpatrick88PRA,Toninelli05PRE,Donati02JNon-Cryst}. We showed
that $\chi_{NL}(t)$ at $T_{d}$ is also a measure of broken ergodicity
- a finding that has been confirmed in computer simulation studies \cite{Thirumalai93PRE}.
A number of studies have also shown that $\chi_{NL}(t)$ also gives
information on the dynamical heterogeneity in glassy liquids \cite{Flenner10PRL,Flenner11PRE,Glotzer00JCP}.

The second transition within
RFOT theory  is a true equilibrium
phase transition occurring at a lower temperature denoted by $T_{K}$,
for historical reasons. The transition at $T_{K}$ is a new type of
equilibrium phase transition, where the order parameter is both random
and, in a technical sense, discontinuous at $T_{K}$, and one that
has a unique and different coherence or correlation length exponent
associated with it compared to other well-studied phase transitions.
%It has recently been emphasized that within the RFOT theory there
%will be yet another transition deep inside the glassy phase. This
%so-called Gardner transition \cite{Kurchan13JPCB} will not be discussed here.

In \cite{Kirkpatrick89PRA} (see also \cite{Kirkpatrick87PRBPotts}) it was argued that in realistic, non-mean-field, systems the transport for $T<T_d$ is activated and that the driving force is entropic. The eventual freezing at the ideal glass transition occurs because the entropic driving force vanishes at $T_K$. This and various scaling arguments were the final ideas in the original \cite{Kirkpatrick89PRA} RFOT theory.

More recently \cite{Maimbourg2016,Kurchan2016,Parisi2020}, there has been an enormous amount of work on liquid systems that can be exactly solved in the high-dimension limit ($d\rightarrow\infty$) and that undergo a SGT. As in earlier work, these theories find two transitions; (i) an ergodicity breaking dynamical transition and (ii) a true equilibrium ideal glass transition. The universal features of these transitions are the same as in the original RFOT of the SGT. It is interesting, and possibly relevant, to note that the fundamental variables in these theories are all related to particle displacements, and not directly to density fluctuations (see however \cite{Szamel2022}).

In recent years there has also been a lot of work on the thermodynamics of metastable or non-equilibrium states within RFOT. This is motivated by the fact that on a restricted time scale, that gets longer and longer further into the glassy phase, these states are almost like equilibrium states. This enlarges the parts of phase space that can be probed by theoretically compressing a fluid.  Among other things, this has led to a connection between the glass transition and the so-called jamming transition \cite{Charbonneau17AnnRevCondMatPhys,Parisi2020}. This work has emphasized that the existence of a Kauzmann temperature is not central to what is referred to as the Gardner transition \cite{Gardner95NucPhysB,Kurchan13JPCB} and the subsequent jamming transition.

The importance of $T_d$ was already anticipated by Goldstein \cite{Goldstein69JCP} (see also \cite{Goldstein2010}), who argued that below a certain temperature transport in liquids is determined by activated transitions. Since then there is overwhelming evidence from 
theoretical, simulation \cite{Thirumalai93PRE}, and experimental \cite{Novikov03PRE} studies that there is a material-dependent
special temperature, $T_{d}$, or temperature region, below which ergodicity breaking, dynamic heterogeneity, aging effects play an important role. This temperature, $T_{d},$ is generally well
above what is conventionally called the glass transition temperature
$T_{g}$, defined roughly when the viscosity reaches a value of
$10^{13}$ poise. The existence of an ideal glass transition at $T_{K}$ in RFOT is
perhaps debatable. It is difficult to unequivocally probe the existence of $T_K$  experimentally because below $T_g$ ($> T_K$) the equilibration times vastly exceed
the observation times. Here, we assume that it does exist, and given that it does,
we determine some of the universal properties associated with this transition.

%In this paper, a number of things are accomplished. We compare and contrast
%both the transitions in the RFOT with the conventional first order  and with second order, or continuous, phase transitions.
%Our aim is to sharpen the differences between these classes of phase
%transitions, so that computer simulations and/or real experiments
%can distinguish between them for the very important problem of the SGT.

The contents of this paper are as follows. In Section II
we discuss the important characteristics of a RFOT, with an eye towards
the DFT of the SGT and on the two distinct (dynamical
and equilibrium) transitions within RFOT. Using scaling and RG ideas we obtain previously unnoticed scaling relations and scaling predictions for the SGT. Activated dynamical scaling ideas are used to predict a number
of effects that should occur as $T_{K}$ is approached. Important
finite size corrections to the ideal glass transition temperature
are also discussed.
In Section III, we compare a RFOT transition
with more conventional phase transition theories for the structural
glass transition. In this section we also make a general argument for a correlation exponent inequality. We conclude in Section IV with a discussion.

\bigskip{}

\section{ Characteristics of the RFOT theory of the SGT}
\label{sec:II}

\subsection*{A. The order parameter and state degeneracy}

The glassy state may be visualized as a frozen liquid state with elastic
properties. To describe a glassy state, we introduce two key ideas.
First, we imagine an order parameter description in terms of frozen
density fluctuations, $\delta n=n-n_{l}$, where $n$ is a local number
density, and $n_{l}$ is the spatially average density that is identical
to the liquid state density. Other order parameters can be imagined,
but frozen density fluctuations are the simplest, and are most directly
related to an important characteristic of a solid: elastic properties
and a nonzero Debye-Waller factor. Because the glassy phase is amorphous
or has random characteristics, the frozen density order parameter
is specified by a functional probability measure $DP[\delta n]$ \cite{Kirkpatrick89JPhysA,Monasson95PRL,Mezard96JPhysA}.
%The first two moments of this measure are, 
%\begin{equation}
%\overline{\delta n(\mathbf{x})}=\int DP[\delta n]\delta n(\mathbf{x})=\frac{1}{V}\int d\mathbf{x}\delta n(\mathbf{x})=0,
%\end{equation}
 %and, the bulk limit is always implied,
%\begin{equation}
%q\equiv\overline{[\delta n(\mathbf{x})]^{2}}=\int DP[\delta n][\delta n(\mathbf{x})]^{2}=\frac{1}{V}\int d\mathbf{x}[\delta n(\mathbf{x})]^{2}.
%\end{equation}
%The final equalities in these equations assume self-averaging. From
%Eq.(2) it follows that the zero in Eq.(1) is a term of $O(V^{-1/2})$
%$with $V$ the system volume and the bulk limit is always taken.

%\subsection*{B. State degeneracy and two transition temperatures }

The second key idea in the formulation of RFOT is that, in general,
one expects a large number of distinct metastable glassy states as
the dynamics starts to become sluggish. If the number of low free
energy metastable states is large enough, this in turn leads to two
distinct transitions. The physical arguments leading to this conclusion
proceed as follows. We denote a particular glassy state by the label
$s$, with the frozen density in that state, $n_{s}=n_{l}+\delta n_{s}$,
and the free energy in that state being equal to $F_{s}$ {[}the density
fields in Eqs. (1) and (2) should also be labelled by $s${]}. 
The first two moments of the order parameter are, 
\begin{equation}
\overline{\delta n_s(\mathbf{x})}=\int DP[\delta n]\delta n_s(\mathbf{x})=\frac{1}{V}\int d\mathbf{x}\delta n_s(\mathbf{x})=0,
\end{equation}
 and 
\begin{equation}
q_{ss'}=\int DP[\delta n]\delta n_{s}(\mathbf{x})\delta n_{s'}(\mathbf{x})=\frac{1}{V}\int d\mathbf{x}\delta n_{s}(\mathbf{x})\delta n_{s'}(\mathbf{x}).
\end{equation}
The final equalities in these equations assume self-averaging. From
Eq.(2.2) it follows that the zero in Eq.(2.1) is a term of $O(V^{-1/2})$
with $V$ the system volume and the bulk limit is always taken.

Exact
calculation for some infinite range spin-glass models as well as for infinite dimensional liquid models \cite{Parisi2020} show that below
a temperature, which we denote by $T_{d}$, there are an extensive number
(the number of states scales like $\exp[\alpha N]$ for a $N$particle
system) of global statistically similar incongruent metastable glassy
states \cite{Huse87JPhysA}. 
In the RFOT, it is assumed that, in a restricted sense (cf
below), this feature also holds in realistic structural glass systems.
Statistically similar states have the same spatially averaged correlation
functions, and incongruent states have zero overlap, 
\begin{equation}
q_{ss'}=\delta_{ss'}q.
\end{equation}
 Because the states are statistically similar one cannot simply use
an external field to pick out a particular state, as could be done
in a regular Ising model for example. The canonical free energy, $F_{c}$,
is then given by the partition function via 
\begin{equation}
Z=\exp[-\beta F_{c}]=Tr\exp[-\beta H]=\sum_{s}\exp[-\beta F_{s}].
\end{equation}
 In the STG, there are two important cases when $F_{c}$ is not the
physical free energy. First, if the barrier between states is actually
infinite then $F_{c}$ cannot be a physically meaningful free energy.
Second, if the barriers are finite but the experimental time scale
is too short for fluctuations to probe the various states, then it
is also not a physical free energy. These considerations also apply
to cases where there is a multiplicity of long-lived metastable states
in any system.

A component averaged free energy can be defined by 
\begin{equation}
\overline{F}=\sum_{s}P_{s}F_{s}
\end{equation}
 with $P_{s}$ the probability to be in the state $s$, 
\begin{equation}
P_{s}=\frac{1}{Z}\exp[-\beta F_{s}]
\end{equation}
 $F_{c}$ and $\overline{F}$ are related by 
\begin{equation}
F_{c}=\overline{F}+T\sum_{s}P_{s}\ln P_{s}\equiv\overline{F}-TS_{s}.\label{driving}
\end{equation}
 Here, $S_{s}$ is the state entropy (sometimes called the complexity,
$I$), which is bounded from above by what is usually meant by the
configurational entropy, $S_{c}$, in non-mean-field models where
there will be transitions between the various states. In general,
$S_{s}$ is related to the solution degeneracy and is extensive (and
$F_{c}\neq\overline{F}$) if there are an exponentially large number
of states. Note that in infinite range models with a RFOT and a nonzero
$S_{s}$ the physical free energy is $\overline{F}$ because the $S_{s}$
in Eq. (\ref{driving}) is an entropy term which is a measure of parts
of state space not explored in a finite amount of time. Since a physical
entropy should only be associated with accessible configurations,
it follows that $F_{c}$ is not a physically meaningful free energy.

The scenario for the two transitions in the RFOT theory is obtained
as follows. For $T>T_{d}$ transport is not collective, and the topology
of state space is trivial. However, as $T\rightarrow T_{d}^{+}$ the
dynamics slows down because at $T_{d}$ an extensive number of statistically
similar, incongruent globally glassy metastable states emerge. If
activated transport is neglected, these states are infinitely long
lived. The liquid state free energy, $F_{l}$ is lower than the physical
glassy state free energy, $\overline{F}$, but it is equal to the
canonical free energy, $F_{c}$. Because there is a multiplicity of glassy
states, a liquid with unit probability will get stuck in one of the
metastable glassy states at temperatures below $T_{d}$. If activated transport
is ignored, it will remain in that state for all times. For infinite
range models with an RFOT, exact dynamical calculations shows a continuous
slowing down and freezing occurs as $T\rightarrow T_{d}^{+}$. The same result
is also found for some approximate, mean-field like, calculations
of dynamics in realistic liquid state models even in the absence of
quenched randomness \cite{Kirkpatrick89JPhysA}. The transition at $T_{d}$ is also closely related
to the temperature at which mode coupling theory of the glass transition predicts a power law divergence of the relaxation times.

In realistic systems, with particles interacting with short-range
interactions, activated transport does take place at $T<T_d$. The
change from diffusive transport to sluggish dynamics, which occurs
at surprisingly low viscosities, is the reason $T_{d}$ is associated
with a dynamical transition. It is a sharp transition only in infinite
range models, but in general it sets a temperature at which the dynamics
becomes glassy like. In support of this interpretation, whose origins
in retrospect can be traced to the insightful arguments given by Goldstein
\cite{Goldstein69JCP}, there is considerable experimental evidence
that at $T_d$ the nature of transport changes from being diffusive
to activated \cite{Novikov03PRE}. As a result, it is our view, that 
theories (for example \cite{Adam65JCP}) in which the analogue of $T_{d}$ does
not naturally emerge cannot describe the STG.

The driving force for the activated transport in the RFOT theory for
temperatures less than $T_{d}$ is entropic, and is given by the state
or configurational entropy, $S_{s}$, Eq.(\ref{driving}). At a lower
temperature denoted by $T_{K}$, after the so-called Kauzmann temperature,
the configurational entropy vanishes, and transport ceases. In other
words, the second transition at $T_{K}$ is the ideal or equilibrium
glass transition temperature. Fits to viscosity data for an extremely
wide class of materials point to the existence of non-zero $T_{K}$
(for a compilation of experimental data see Ref. \cite{Capaccioli08JPCB}).

\subsection*{B. The dynamic transition and the associated correlation length}

We characterize the dynamical transition as follows. Consider the
dynamical order parameter, 
\begin{equation}
q(\mathbf{x}-\mathbf{y},t_1-t_2)=\langle\hat{q}(\mathbf{x},\mathbf{y},t_1,t_2)\rangle=\langle\delta n(\mathbf{x},t_1)\delta n(\mathbf{y},t_2)\rangle,
\end{equation}
 Above $T_{d}$, this correlation decays as $t\rightarrow\infty$
but as $T\rightarrow T_{d}^{+}$ the decay gets slower and slower
in a power law fashion. At $T_{d}$ it no longer decays, (except in
non-mean-field models) on the longest time scale. Effectively, it
is the Edwards-Anderson order parameter for the glass transition:
\begin{equation}
q_{EA}=\lim_{t\rightarrow\infty}q(\mathbf{0},t).
\end{equation}
Because the order parameter involves the square of the density fluctuations
it is clear that the susceptibility will be non-linear involving higher
order correlations \cite{Kirkpatrick88PRA}, 
\begin{equation}
\begin{split}
&\chi_{NL}(\mathbf{x}-\mathbf{y},t_1-t_3,t_4-t_2,t=t_4-t_1)=\\
&\langle\hat{q}(\mathbf{x},\mathbf{y},t_1,t_2)\hat{q}(\mathbf{y},\mathbf{x},t_3,t_4)\rangle_c.
\end{split}
\end{equation}
Where $\langle \rangle_c$ denotes cumulant average where all pairwise density correlations are subtracted out. In an exactly soluble spin-glass model we have previously shown \cite{Kirkpatrick88PRA} that the 'static' susceptibility for the
glass transition at $T_{d}$ is 
\begin{equation}
\begin{split}
&\chi_{NL}(k,t<\tau_c)=\\
&\int d(t_1-t_3)d(t_4-t_2)d(\mathbf{x}-\mathbf{y})\exp(-i\mathbf{k}\cdot(\mathbf{x}-\mathbf{y}))\\
&\chi_{NL}(\mathbf{x}-\mathbf{y},t_1-t_3,t_4-t_2,t=t_4-t_1<\tau_c),
\end{split}
\end{equation}
where $\tau_c$ is the time that the plateau exist in $q(t)$ (it diverges at $T_d$ in the mean-field limit).
%It is also interesting to define a dynamical frequency dependent
%susceptibility using, 
%\begin{equation}
%\chi_{NL}(k,\omega)=\int dt\int d\mathbf{x}\exp(-i\mathbf{k\cdot x}+i\omega t)\chi_{NL}(\mathbf{x},t).
%\end{equation}
As $r=T/T_{d}-1\rightarrow0^+$, the homogeneous static susceptibility,
$\chi_{NL}=\chi_{NL}(k\rightarrow\mathbf{0})$, in a mean-field like
theory, diverges as \cite{Kirkpatrick88PRA}, 
\begin{equation}
\chi_{NL}\sim1/\sqrt{r},
\end{equation}
and at finite and small wavenumber, 
\begin{equation}
\chi_{NL}(\mathbf{\mathrm{k})}\sim1/[k^{2}+\xi_{o}^{-2}\sqrt{r}]
\end{equation}
 with $\xi_{o}$ a microscopic (correlation) length. This in turn
defines a divergent length scale as $T\rightarrow T_{d}^{+}$ given
by, 
\begin{equation}
\xi\sim\xi_{o}/r^{1/4}.
\end{equation}
 This is the same divergence seen at a mean-field spinodal point. These same results with $r\rightarrow|r|$ were obtained \cite{Kirkpatrick87PRBPotts} in an exactly soluble spin-glass model for
 $r=T/T_{d}-1\rightarrow0^-$.

 The limiting procedure implied by Eq.(2.11) is crucial to obtain the correct diverging non-linear susceptibility result \cite{Kirkpatrick88PRA}. In \cite{Kirkpatrick88PRA} it was suggested that this is a reasonable definition of an ergodic to non-ergodic transition.

%There is considerable on going work beyond mean-field theory to describe
%the dynamics in glassy liquids just right below $T_{d}$. Remarkably,
%the dynamics has been related to a cubic field theory with a random
%field \cite{Franz11EPJE, Biroli13}. This in turn means that ideas and technical results for the
%magnetic random field problems (which do not typically have a cubic
%term) can be applied. A very interesting aspect of these works is that
%it suggest that somewhere in the phase diagram there is a critical
%point with properties related to the random field Ising model. It
%has even been suggested that the experimental glass transition at
%$T_{g}$ is generically due to this critical point being in proximity.

\subsection*{C. The equilibrium transition and its correlation length}

According to RFOT, the ideal glass transition at $T_{K}$ is a discontinuous
transition. Hence, care must be taken in defining and interpreting
a correlation or coherence length. For ordinary first order phase
transitions (liquid to a periodic crystal for example) Fisher and
Berker (FB) \cite{Fisher84PRB}  have formally described the meaning of such a length scale.
They conclude that there are two distinct interpretations of a divergent
length, referred to as coherence or persistent length for a first
order transition. The first, using a scaling analysis and the observation
that the definition of long range order is consistent with a correlation
length exponent given by $\nu_{H}=1/d$, with $d$ being the spatial
dimension. The second is via a finite-size scaling analysis and the
divergent length is the length scale where the sharp first order phase
transition becomes rounded in a finite system. This correlation length
exponent is also given by $\tilde{\nu}=1/d$.

Here, we generalize the arguments of FB to a RFOT, further elaborating
on our previous study \cite{Kirkpatrick89PRA}. We use the
density order parameter discussed in Section II.A. For the conjugate
field we use a chemical potential $h=\mu(T,p)-\mu(T_{K},p_{K})$,
allowing for the location of the glass transition to depend on both
temperature and pressure.

We then give two further distinct arguments for the exponent characterizing
the divergence of the correlation length. The first is an exact calculation
for an infinite range spin-glass model with a RFOT that gives an
identical result. The final argument that gives the same exponent
is a partly dynamical/nucleation-like one and is presented in Sec
II.E. All together there are four separate arguments that give the
same result.

\subsubsection*{1. Scaling analysis}

If we define a renormalization group (RG) length rescaling factor
by $b$ then the conjugate field $h$ should be associated with the
largest RG eigenvalue, $b^{\lambda}$, and therefore should scale
as, 
\begin{equation}
h'=b^{\lambda}h
\end{equation}
 while the free energy density will scale like, 
\begin{equation}
f(h)=b^{-d}f(h').
\end{equation}
 In addition, the scaling of the correlation length should by definition
follow, 
\begin{equation}
\xi(h)=b\xi(h').
\end{equation}
 If we choose $b=h^{-1/\lambda}$ then we formally obtain, 
\begin{equation}
f(h)=h^{d/\lambda}f(1),
\end{equation}
 and
 \begin{equation}
\xi(h)=h^{-1/\lambda}\xi(1).
\label{corr1}
\end{equation}
 Now, the order parameter $q$ is related to the second derivative
of $f$ with respect to $h$ or, 
\begin{equation}
q\sim h^{d/\lambda-2}.
\end{equation}
 At the RFOT, $q$ is discontinuous which gives $\lambda=d/2$. Equation
(\ref{corr1}), on the other hand implies that the correlation length
exponent is $\nu_{h}=1/\lambda$, or 
\begin{equation}
\nu_{h}=2/d.
\end{equation}
 These considerations establish, rather rigorously, that as long as
long as FB type scaling holds near a first order transition it is
inevitable that the correlation length should diverge at $T_{K}$
with an exponent $\frac{2}{d}$ at RFOT. The difference in the values
of the correlation length exponent is due to the differing nature
of the order parameter describing RFOT and first order transitions.
In the first order transition density itself is discontinuous whereas in
RFOT at $T_{K}$ it is the square of the density (EA parameter) that
changes abruptly.

\subsubsection*{2. Finite size effects}

Next we consider a system with finite size, $L$, and do a finite
size scaling argument. We postulate that the scaling part of the free
energy behaves as, 
\begin{equation}
f(h,L)\sim L^{-\zeta}f(L/\widetilde{\xi})\equiv L^{-\zeta}Y(hL^{1/\widetilde{\nu}})
\end{equation}
 where $\xi\sim h^{-\widetilde{\nu}}$ is the (finite-size correlation
or coherence length) length scale where finite size rounding of the
RFOT takes place at $T_{K}$. By noting that the order parameter $q$
is related to the second derivative of $f$ with respect to $h$,
we find, 
\begin{equation}
q\sim L^{2/\widetilde{\nu}-\zeta}Y''(hL^{1/\widetilde{\nu}})
\end{equation}
 The discontinuous nature of the transition gives, 
\begin{equation}
\zeta=2/\widetilde{\nu}
\end{equation}
 Note that if we used dimensional analysis to obtain $\zeta=d$ then
it immediately follows that 
\begin{equation}
\widetilde{\nu}=2/d
\end{equation}

As an alternative, we follow the FB treatment, and examine the susceptibility,
$\chi$, for this transition. Because the order parameter for the
transition already involves two density fluctuations, the susceptibility,
as used in Section I.C, will involve four density fields. Accordingly,
it is related to the fourth derivative of $f$ with respect to $h$.
The finite size scaling relation will then be something like, 
\begin{equation}
\chi(h,L)\sim L^{2/\widetilde{\nu}}Y''''(hL^{1/\widetilde{\nu}})
\end{equation}
 One can then argue that $\chi$ will attain a maximum value at $h=0$,
and will scale as $\sim L^{d}$. This in turn again gives $\widetilde{\nu}=2/d$.
The inescapable conclusion is that the arguments leading to $\widetilde{\nu}=2/d$
are robust, and theoretically well-founded.

\subsubsection*{3. Exact results for an infinite range model}

Finally, there is an exact calculation for an infinite range p-state
($p>4)$ Potts spin glass model with a RFOT \cite{Kirkpatrick87PRBPotts,Kirkpatrick88PRBPotts}
that gives a length scale that diverges with a correlation length
exponent of $\nu=2/d$ as the transition is approached. In particular,
it has been shown that as $T_{K}$ is approached from above the complexity
or state entropy (which here and below we denote by the configurational
entropy, $S_{c}$) behaves as, 
\begin{equation}
S_{c}\sim\exp[N\epsilon]
\end{equation}
 with $N$ the number of lattice sites ($N\rightarrow\infty$) and
$\epsilon=1-T/T_{K}$ the dimensionless distance from the RFOT. On
the other hand, as $T_{K}$ is approached from below one finds \cite{Kirkpatrick87PRBPotts} that
$S_{c}$ diverges as, 
\begin{equation}
S_{c}\sim\exp[1/|\epsilon|]
\end{equation}
 If we identify $N$ with a system volume $\sim L^{d}$ then these
two equations give a length scale where even below $T_{K}$ the system
behaves as though it was not frozen. Defining a correlation length,
$\xi$, in this way gives, 
\begin{equation}
\xi\sim1/|\epsilon|^{2/d}
\end{equation}
 or, 
\begin{equation}
\nu=2/d.
\end{equation}
 Notice that this definition of a correlation length is closely related
to a finite size correlation length.

\subsubsection*{4. Finite size shift in the glass transition temperature}

Understanding glassy physics in a finite geometry is technologically
important, and is needed to analyze results from computer simulations.
From a phase transition perspective, many authors starting with the pioneering studies by Ferdinand and Fisher \cite{Ferdinand69PR}  have stressed that
there are two important finite size effects at any phase transition.
The first is the rounding of the transition as described in Sec II.D.2
above. The second distinct effect is the finite size shift in the
transition temperature. Naturally, we expect them both to be relevant
near RFOTs.

Physically, the $\epsilon$ (or $h$) in the above expressions should
be replaced by $\dot{\epsilon}=\epsilon-\Delta\epsilon(L)$, where
$\Delta\epsilon(L)$ measures the shift in the glass transition temperature
due to the finite size. We give two distinct arguments for the $L$
dependence of $\Delta\epsilon(L)$. The first uses that the free energy
difference between two coexisting states at the RFOT is of order $\sim T_{K}L^{d/2}$.
The temperature difference, $\Delta T$, where the system cannot distinguish
between the two states is the total free energy difference, which
is proportional to $\sim L^{d}\Delta T$. Equating these two free
energies we obtain, 
\begin{equation}
\Delta\epsilon(L)\sim\frac{1}{L^{d/2}}\label{LdepT}
\end{equation}

The second argument hinges on an old result of Imry \cite{Imry80PRB}
who suggested that at an ordinary first order phase transition the
finite size shift in the transition temperature is given by, 
\begin{equation}
\Delta T\sim\frac{1}{L^{d}s}
\end{equation}
 where $s$ is the latent heat associated with the transition. At
an RFOT, though, the latent heat vanishes as the configurational entropy
or, $s\sim\xi^{-1/\nu}\sim L^{-d/2}$ in a finite system. Using this
we recover the first argument for $\Delta\epsilon(L)$.

If the predicted $L$-dependent shift in the ideal glass transition temperature can
be systematically studied then measuring the above exponent would
be an important verification of the RFOT theory. Recent important
developments in computer simulations probing the equilibrium glass
transition by random pinning of a fraction of particles \cite{Berthier12PRE} as $L$ is
varied could be a way of verifying Eq. ({\ref{LdepT}).

\subsection*{D. The specific heat at the equilibrium RFOT transition}

First order phase transitions are associated with strong coupling
RG fixed points \cite{Fisher84PRB}. As a result, dangerous irrelevant variables are not
expected to play any role, which in turn implies that hyperscaling
is valid in all dimensions. If the same applies to random first order
phase transitions, then the standard hyperscaling relation for the
specific heat exponent, 
\begin{equation}
\alpha=2-\nu d
\end{equation}
 should always hold. With $\nu=2/d$ one then finds that $\alpha=0$
at a RFOT. This scaling result for the specific heat is also consistent
with exact calculations for infinite range models with an RFOT. For
structural glasses, this result presumably implies a discontinuity
in the specific heat, as is always observed experimentally. 

It is worth emphasizing at this juncture that the Adams-Gibbs theory
would predicts $\alpha=1$ (a latent heat) if hyperscaling is assumed
because in the AG theory $\nu=1/d$.

\subsection*{E. Transport near the equilibrium RFOT}

An important characteristic of a glass transition is the occurrence
of extremely long time scales associated with transport below $T_{d}$.
The critical slowing down at an ordinary transition means that the
time scale grows as a power of the correlation length, $\tau\sim\xi^{z}$
with $z$ being the dynamical scaling exponent. In sharp contrast,
at a glass transition the critical time scale grows exponentially
with $\xi$, 
\begin{equation}
\ln(\tau/\tau_m)\sim\xi^{\psi}\label{activated}
\end{equation}
 with $\tau_m$ a microscopic time scale, and $\psi$ a generalized
dynamical scaling exponent. Effectively, this implies $z=\infty$.
As a result of such extreme slowing down, the equilibrium behavior
of the system near $T_{K}$ becomes inaccessible
for all practical purposes. In other words, the experimentally accessible
time scales are not sufficient for the system to reach equilibrium.
It is in this sense the glassy system falls out of equilibrium.

Activated scaling, as described by Eq. \ref{activated}, follows from
a barrier crossing picture of the system's free energy landscape.
In the context of the STG, it is related to cooperative motions \cite{Adam65JCP}
involving particles within a length scale that continues to grow as
$T\rightarrow T_{K}$. In the temperature range, $T_{d}>T>T_{K}$,
there is an entropic driving force that causes a, compact, glassy
state of size $\xi^{d}$ to make a transition to a different glassy
state, with approximately the same free energy, also of size $\xi^{d}$.
The physical picture that results is a system that looks like a patchwork,
of different glassy regions separated by diffuse or fuzzy interfaces
slowly making a transition to yet other glassy states. This is also
called the mosaic state. For the uncorrelated states that exist above
$T_{K}$, the law of large numbers is consistent with a barrier that
scale like $\sim\xi^{d/2}$. This is also consistent with scaling
and an entropic driving force $\epsilon\xi^{d}\sim\xi^{d-1/\nu}$,
if $\nu=2/d$. {Thus}, the driving force and the free energy barrier
scale in the same way, if and only if $\nu=2/d$. Activated transport
with the barriers scaling as $\sim\xi^{d/2}$ was also recently argued
by Langer \cite{Langer13PRE}.

In the original RFOT paper, we modified a wetting argument first given by Villain \cite{Villain85JPhysique}
for the random field Ising model and applied it to the structural
glass problem which also led to barriers scaling like $\xi^{d/2}$.
Here, we give a different but related argument. We first imagine a
putative compact glassy state of size $L^{d}$. There will in general
two distinct driving forces that can cause it to make a transition
to another glassy state: An entropic one scaling like $\sim\epsilon L^{d}$
and a free energy one $\sim L^{d/2}$ because there are states that
typically have a lower free energy by this amount. Opposing the transition
is a barrier that naively scales like a surface tension term $\sim\sigma L^{d-1}$.
Now assume a bump (or an excitation) of scale also $\sim L$ (another
scale can be introduced but the final result is not altered) forms
on the interface because of the free energy driving force. The free
energy change due to the creation of the bump is, 
\begin{equation}
\delta F\sim-L^{d/2}+\sigma L^{d-1}
\end{equation}
 The physical effect of the bump forming is that the surface tension
is reduced compared to its naive value. The scale dependent surface
tension can be obtained by minimizing $\delta F$ with respect to
$r$ with the result, 
\begin{equation}
\sigma(L)\sim L^{-(d-2)/2}
\end{equation}
 The barriers to transport therefore scale as $\sim\sigma(L)L^{d-1}=L^{d/2}$.

All of these arguments in turn imply a Vogel-Fulcher law for relaxation
as the glass transition is approached, 
\begin{equation}
\tau\sim\tau_m\exp[\frac{D}{T/T_{K}-1}]
\label{relax}
\end{equation}
 with $D$ a positive constant.

Note that the barriers scaling like $\xi^{d/2}$ implies there is
really no interface between two of the statistical similar glassy
states. Rather, to go from one state to another, roughly $N^{1/2}$
of the $N$ particles in a correlated volume must be rearranged, as
noted earlier by Bouchaud and Biroli \cite{Bouchaud04JCP}. 

\subsection*{F. Rare regions near the glass transition}

The existence of DH suggests that in a very viscous liquid
the longest time decay of any time correlation function will be
determined by the large rare region or anomalous clusters of
particles of some linear dimension $L$ \cite{KirkpatrickRMP2015}. These
large clusters are fluidized and can relax to a more typical
configuration of particles in some characteristic time $\tau(L)$. For
this argument to be sensible$L$ must be larger than a molecular
scale. To estimate the effect of these large rare regions on a
typical time correlation function an average over $L$ must be
performed.

Since the large clusters are rare, we assume that their
probability distribution is controlled by Poisson statistics so
that the tail probability of an unusual cluster of size $L$ is
\begin{equation}
P(L)\propto\exp(-cL^d).
\end{equation}
We consider a diffusive correlation function, $C(k,t)\propto\exp(-Dk^2t)$, and we assume the relevant  wavenumber scales as $k\propto 1/L$, and a scale dependent diffusion coefficient $D(L)$. That is, we consider,
\begin{equation}
C(L,t)\propto\exp(-t/\tau_D(L)),
\end{equation}
with $\tau_D(L)=L^2/D(L)$ the diffusive time scale.
Since $D(L)$ has units of $length^2/time$ we take $\tau_D$ to be proportional to the RFOT time scale,
\begin{equation}
\tau_D(L)\propto\tau_{RFOT}(L)=\tau_m\exp(aL^{d/2}),
\end{equation}
with $\tau_m$ a microscopic time and $a$ a positive constant.

Using all this the average correlation function deep in the supercooled region then decays for long times as  \cite{KirkpatrickRMP2015},
\begin{equation}
\begin{split}
C(t)\propto\int dL P(L) \exp(-(t/\tau_m)e^{-aL^{d/2}})\\
\propto\exp[-A(\ln t/\tau_m)^2],
\end{split}
\end{equation}
with $A$ a positive constant. We conclude that for long times $C(t)$ decays faster than any power law, but slower than any exponential. Note that the saddle point evaluation gives the characteristic length scale $L(t)\propto (\ln t)^{2/d}$.

Finally if we define a distribution of relaxation times, $P(\tau)$, by
\begin{equation}
C(t)\propto\int d\tau P(\tau)\exp(-t/\tau),
\end{equation}
then,
\begin{equation}
P(\tau\rightarrow\infty)\propto\exp[-c(\ln t/\tau_m)^2],
\end{equation}
again, slower than any exponential.

\subsection*{G. Activated scaling as $\mathbf{T}$ approaches $\mathbf{T_{K}}$}

Activated scaling was developed to understand finite dimensional (i.,e.,
three dimensions) spin glasses and random field magnets where the
dynamics is controlled by large, possibly divergent, free energy barriers \cite{Fisher88PRB}.
Similar ideas can be applied to the structural glass problem, also
in three-dimensions.

Here we examine the behavior of the glass transition susceptibility,
introduced in Sec II.C using activated scaling ideas \cite{Fisher88PRB} as the ideal
glass transition is approached. First we define a slightly different non-linear susceptibility (see Eqs.(2.10) and (2.11)),
\begin{equation}
\begin{split}
&\tilde{\chi}_{NL}(k,t)=\int d(\mathbf{x}-\mathbf{y})\exp(-i\mathbf{k}\cdot(\mathbf{x}-\mathbf{y}))\\
&\chi_{NL}(\mathbf{x}-\mathbf{y},t,t,0),
\end{split}
\end{equation}
We then start with the observation, discussed
in SecII.D, that the scale dimension of $q(\mathbf{x},t)$ is zero.
This and the activated scaling ansatz implies that the wavenumber
and time dependent glass transition susceptibility will satisfy the
scaling law  \cite{KirkpatrickRMP2015}, 
\begin{equation}
\tilde{\chi}_{NL}(\epsilon,k,t)=b^{d}F_{\chi}[\epsilon b^{1/\nu},bk,\frac{b^{d/2}}{\ln(t/\tau_m)}]\label{suseq}
\end{equation}
where $\epsilon=T/T_{K}-1$ is the dimensionless distance from the
ideal glass transition, $\tau_m$ is some microscopic time scale, and
$F_{\chi}$ is a scaling function. Note that we have used here that
the barrier height scales as $b^{d/2}\sim\xi^{d/2}$. This equation
implies a number of non-trivial results. For example, at zero wavenumber,
and at the ideal glass transition temperature we can choose $b=[\ln(t/\tau_m)]^{2/d}$
to obtain, 
\begin{equation}
\tilde{\chi}_{NL}(0,0,t\rightarrow\infty)\sim[\ln(t/\tau_m)]^{2}.\label{susNL}
\end{equation}
 This dynamic scaling result is valid as long as $\epsilon\ln(t/\tau_m)<1$.
This in also defines a dynamic crossover $\epsilon$ being given by,
\begin{equation}
\epsilon_{x}\sim1/[\ln(t/\tau_m)]\label{crossover}
\end{equation}
 Physically this means that the large correlations that exist at $T_{K}$
can be measured by examining the slow growth in time of the glass
transition susceptibility around $k=0$. This should be experimentally
relevant. If the exponent of $2$ in Eq.(\ref{susNL}) can be experimentally
demonstrated then that would be very strong evidence for the validity
of the RFOT theory of the SGT.

The frequency dependent glass transition susceptibility defined by
Eq.(\ref{susNL}) can similarly be expressed as a scaling function.
In general the $\epsilon_{x}$ given by Eq.(\ref{crossover}) will
give the scale distinguishing static critical behavior from dynamical
critical behavior for all quantities as $T\rightarrow T_{K}$.

Although not as rigorously founded as the scaling law for $\tilde{\chi}_{NL}$,
we can also give a scaling law for the frequency dependent shear viscosity,
$\eta(\epsilon,\omega)$, if we assume that because it is related
to a time integral of a time correlation function its static value
is proportional to $\tau$ given by Eq. (38). We then obtain  \cite{KirkpatrickRMP2015},
\begin{equation}
\eta(\epsilon,\omega)=\exp(b^{d/2})F_{\eta}[\epsilon b^{1/\nu},\frac{b^{d/2}}{\ln(1/\tau_m\omega)}]
\label{eta}
\end{equation}
with $F_{\eta}$ a scaling function. The static or zero frequency
shear viscosity then behaves as $\tau$ but for $\epsilon<1/\ln(1/\tau_m\omega)$
it behaves as  \cite{KirkpatrickRMP2015}
\begin{equation}
\eta[\epsilon\ln(1/\tau_m\omega)<1]\sim\frac{1}{\tau_m\omega}.
\label{etaglassy}
\end{equation}
Again, the important physical and experimental point is that $\epsilon_{x}$
given by Eq.( \ref{crossover}) sets the crossover scale in either time
or frequency ($t\rightarrow1/\omega$) space. Note that $\eta$ being simply proportional to $\tau$ in Eq.(\ref{eta}) is needed to obtain Eq.(\ref{etaglassy}), which in turn is required for the proper stress/strain relation in the glassy phase. Also note that Eq.(2.49) in the Naiver-Stokes equation will lead to shear waves in this frequency range, which is certainly the correct physical result for frequencies not too small near the glass transition. The crucial result from activated scaling is that $\epsilon<1/\ln(1/\tau_m\omega)$ sets the frequency scale where shear waves should be observed in the supercooled liquid phase.

%\subsection*{\bigskip{}}

\section{Comparision to other phase transition theories for the
glass transition and an inequality}
\label{sec:III}

In this Section, we  compare  and contrast the
RFOT theory with conventional first and second order phase transitions, with particular focus on the
SGT. 

\subsection*{A. First order phase transition theory for the glass transition}

An earlier DFT of the SGT, advanced in the in the mid-eighties \cite{Singh85PRL,Dasgupta99PRE}, was based on  the approximate Ramakrishnan-Yussouff \cite{Ramakrishnan79PRB} theory of the liquid to crystal transition and modified it for the SGT. 
By generating random sites of a large
system to describe the amorphous state (viewed as an aperiodic solid
instead of periodic crystal characterizing an ordered solid) it was
shown numerically that the RY free energy functional has a metastable
solution, which presumably corresponds to a static description of
the glassy state. The transition to such an aperiodic state was deemed
to be first order, occurring at a density below the putative crystallization
density.

Although such a description might capture aspects of the amorphous state it differs qualitatively from the RFOT in two important aspects: (1) The transition to the aperiodic
state is first order with density itself exhibiting a discontinuity
where as in RFOT it is the analogue of the EA parameter describing density fluctuations (Eq. 2) that has a
jump at $T_{K}$.  A regular first order transition would also have
a latent heat, and exhibit something that resembles a discontinuity
in the specific heat. In contrast, RFOT theory predicts absence of latent heat at $T_K$. It is also unclear if the identified freezing density is close to the analogue of $T_{K}$. (2) The RY theory was not used
to identify a dynamical transition temperature or density. More importantly,
the link between $T_{d}$ and $T_{K}$ requiring the emergence of
exponentially large number of low free energy metastable glassy states,
which is the most important aspect of the RFOT theory was not demonstrated.
(3) Finally, if one were to associate a divergent length scale or
exponent with an ordinary first order phase transition then it would
be the Berker-Fisher value $\nu=1/d$ discussed in Sec II.D. Such
a value is inconsistent with the equality found using the RFOT theory (further discussed in Sec
III.C), nor is it in accord with recent simulation results \cite{Mosayebi10PRL,Tanaka10NATMAT}.

\subsection*{B. Second order phase transition theory for the SGT }

Over the years there have been numerous theories
of the SGT that can be classified as standard second order or continuous
phase transitions. The earliest ones were the Adam-Gibbs (AG) \cite{Adam65JCP} and Gibbs-DiMarzio
(GDM) and DiMarzio of the SGT and rubber transitions in polymer systems \cite{Gibbs58JCP}.
A scenario along these lines has been postulated by Tanaka \cite{Tanaka10NATMAT,Tanaka2022} based on computer simulations, which has been  further discussed by Langer \cite{Langer13PRE}. It is worth remarking that these theories do not consider consequences of the change in the nature of transport at $T \approx T_d$.

We first remark that it is hard to imagine how any equilibrium
order parameter description of a liquid to glass transition would
not generically have a discontinuous character, at least according to
Landau theory. At isolated critical points, a continuous transition 
is possible. In fact, within RFOT this has been discussed by others,
and related to a critical point for a random field problem \cite{Franz11EPJE}. Second,
in a second order transition the concept of a metastable state is
not obvious. As a consequence it is difficult to see how such transitions
can have a temperature crossover scale (in RFOT, it is $T_{d}$) and
also a a true equilibrium transition temperature (in RFOT, it is $T_{K}$)
at lower temperatures. The existence of two distinct temperatures
in supercooled liquids seems to be well confirmed both numerically,
as well as experimentally. The existence of the temperature scale
$T_{d}$ at which loss of ergodicity starts to manifest itself \cite{Thirumalai89PRA} also tidily explains dynamic heterogeneity
and aging. Technically, the problem with a second order phase transition
scenario is that by definition there is only one temperature where the
system becomes unstable to infinitesimal fluctuations everywhere.

\subsection*{C. Self-generated quenched disorder implies $\boldsymbol{\nu\geq2/d}$}

\textcolor{black}{For systems with quenched disorder, the correlation
length exponent at any phase transition must satisfy the inequality
$\nu\geq2/d$. For the structural glass problem, there is effectively
quenched disorder that is self generated. Here, we argue the self
generated quenched randomness in the structural glass problem also
leads to the inequality $\nu\geq2/d$.}

We begin by noting that below $T_{d}$ (technically
close to $T_{K}$) the glassy state can be partitioned into cluster
of particles of size $\xi$. Because the observation time is much
greater than the relaxation time for particles in two clusters to
interchange, the motion of particles in one cluster is essentially
isolated from another. Since $\xi$ is sufficiently large each cooperatively
rearranging (CRR) region will have differing thermodynamic properties,
which clearly is a violation of large numbers. In particular, we expect
fluctuations of the ideal glass transition temperature, $\delta T_{K}$,
in the various clusters to scale as $\sqrt{(\delta T_{K})^{2}}\sim1/\xi^{d/2}$.
In order for the ideal glass transition to exist, this fluctuation
must be smaller than the distance to the ideal glass transition, $\epsilon\sim1/\xi^{1/\nu}$.
Consequently, $\nu$ must satisfy $\nu\geq2/d$.

\section{ Discussion}
\label{sec:IV}

%In this paper we have attempted to clarify, using
%a variety of arguments, the differences between RFOT and conventional first and second order phase transitions. 
The RFOT theory provides a coherent
picture of the liquid state as it is cooled and supercooled. We have given physical arguments,
rooted in theory, that there are two relevant temperatures, one attributable to change
in transport at $T_{d}$ and the other in which a genuine equilibrium
transition occurs at $T_K$ with both first and second order characteristics. Both these temperatures emerge
from a single static theory. By expanding on our prior works \cite{Kirkpatrick87PRL,Kirkpatrick88PRA,Kirkpatrick89PRA} we obtained a number of results. (1) We give a variety of arguments
to show that a characteristic length diverges at $T_{K}$ with an
exponent $\nu=2/d$. In many ways this makes rigorous our earlier
predictions \cite{Kirkpatrick89PRA}, and this finding is an important corner stone of RFOT. (2) We 
predict that there are large finite size effects at $T_{K}$.
The theory predicts a specific rounding of the transition, as well
as a specific shift in the location of $T_{K}$, due to finite size
effects. This implies that there will be variations in the value of
$T_{K}$ depending on the system size. This prediction is amenable to test by computer
simulations using the random pinning method \cite{Berthier12PRE,Kob13PRL,Karmakar13PNAS}. (3) Using ideas based on activated scaling \cite{Fisher88PRB}, we predict that the
non-linear susceptibility near $T_{K}$ on long length scales diverges
as $\sim[ln(t/\tau_m)]^{2}$. We have also obtained the cross-over
scale that distinguishes between dynamical critical effects and static
critical effects as $T_{K}$ is approached. This is especially relevant for the shear viscosity.

The equilibrium transition in the RFOT theory is a novel type of transition
and as such, it has certain universal properties. The implication
is that most of the features of the ideal glass transition discussed
here are independent of the Kauzmann-like interpretation of $T_{K}$.
For example, the order parameter definition given by Eqs. (1) and
(2) does not hinge on an entropy or state solution like crisis, rather
it is motivated by a glass being characterized by a frozen and random
density order parameter. Physically this seems necessary to obtain
the elastic behavior of a glass. Any field theory, or Landau theory,
for such a transition will necessarily have a first order nature because general symmetry allows for  a cubic term in the theory. This
implies in general, that if an ideal glass transition exists with such
an order parameter it will be a RFOT with some sort of divergent
length scale, characterized by an exponent of $\nu=2/d$. Given this,
results like the specific heat behavior discussed in Section II.E
will be equally generic. The inequality $\nu\ge 2/d$ also seems generic. To discuss transport near an ideal glass
transition we have largely used many state type arguments to obtain
free energy barriers that scale like $\xi^{d/2}$. However, this result
too seems to go beyond an entropy crisis picture: It only depends
on local particle rearrangements to go from two uncorrelated one glassy configurations.
The law of large numbers suggests that
if the volume in question contains $N$-particles then this sort of
fluctuation will scale as $N^{1/2}\sim\xi^{d/2}$ \cite{Kirkpatrick89JPhysA,Thirumalai95JPI}. The fundamental
conclusion from this sort of reasoning is that results like Eq. (\ref{relax}), Eq. (\ref{suseq}) and Eq. (\ref{susNL}) should hold even if mechanism for an ideal glass transition
is not precisely a Kauzmann-like transition.

In deriving the universal aspects of RFOT at $T_K$ we have assumed that the equilibrium ideal glass transition exist. The temperature dependence of viscosity of a large class of glass forming materials have been analyzed using the VFT law with non-zero $T_K$ \cite{Capaccioli08JPCB}. More recently, computer simulations point to the existence of an equilibrium transition below $T_d$ \cite{Kob13PRL}. Among them the most notable ones are those which have observed a phase transitions in model glass forming systems by randomly pinning a fraction of particles at low temperatures \cite{Berthier12PRE,Kob13PRL,Karmakar13PNAS}. This procedure, which apparently allows one to probe equilibrium behavior, clearly shows that an equilibrium transition with characteristics predicted by the RFOT theory does occur. Although the wealth of experimental data as well computer simulations provides strong evidence that $T_K \ne 0$ additional tests of universal behavior predicted here for RFOT are needed before fully validating all the aspects of the RFOT  theory of the SGT.  

Historically, the Adam-Gibbs (AG) theory \cite{Adam65JCP} was the
first to propose an explanation for the possible divergence of relaxation
times at $T_{K}$ based on vanishing of configurational entropy. The
predictions based on the more \textit{ad hoc}
AG mechanism is distinct from the RFOT theory in at least two ways.
First, in the AG theory there is no analogue of $T_{d}$ giving the
impression that the character of transport changes only at $T_{K}$
or at all temperatures. This would be inconsistent with many observations
suggesting that
at $T_{d}$, particle transport changes character, and theoretically with numerous exact dynamical solutions in high dimensions \cite{Maimbourg2016,Kurchan2016,Parisi2020}. Indeed, there
are innumerable computer simulations that experimentally support this picture. More
importantly, by analyzing data for a number of glass forming materials Novikov and Sokolov
have proposed that there may be a near \textquotedbl{}universal\textquotedbl{}
value of 10$^{-7}$s \cite{Novikov03PRE} (larger than the estimate by Goldstein \cite{Goldstein69JCP}) for relaxation, which sets the boundary between
diffusive transport and onset of activated transitions. Of course,
the value of $T_{d}$ at which relaxation exceeds 10$^{-7}$s is clearly
material dependent. Several computer simulations have noted that the nature of transport (non-Gaussian behavior in van Hove correlation functions and evidence for hopping mechanism \cite{Miyagawa88JCP,Thirumalai93PRE,Kob95PRE}) changes even before $T_d$ is reached, further supporting the accepted view that the dynamical transition is avoided. We surmise that theories, such as the AG theory, which cannot describe the dynamical changes at $T_d$ may not be suitable for describing glass transition.  Second, RFOT predicts a much stronger growth of
the correlation length than the AG theory. Interestingly, 
recent computer simulation studies support the predictions of RFOT
theory first by identifying the existence of $T_{K}$ (or equivalently density \cite{Kang13PRE}), and second
by confirming that $\nu=\frac{2}{d}$ \cite{Mosayebi10PRL,Tanaka10NATMAT}.

Universal aspects of  mode coupling theory have been remarkably successful
in describing the change in the character of relaxation of undercooled
liquids in the neighborhood of $T_{d}$. However, because MCT predicts
that relaxation times only grow as power law as temperature is decreased
it cannot account for activated transitions. As a result MCT cannot
describe transport deep into the supercooled regime.

In our view, currently RFOT theory encompassing
both $T_{d}$ and $T_{K}$, appears to be the only theory that accounts
for changes in phase space structure, break down of effective ergodicity, emergence of dynamic heterogeneity as a 
consequence of violation of law of large numbers,
and growth of $\xi$. Most importantly, both the equilibrium and dynamical
aspects at $T_{d}$ and $T_{K}$ emerge naturally without having to
treat the physics at these temperatures separately as done by MCT
or the AG theory. The result of the RFOT theory is that analysis of experimental relaxation data must be done in two steps. For $T \le T_d$ the relaxation time should be analyzed using a power law. In the temperature range, $T_K \le T \le T_d$ the temperature dependence of the relaxation data should follow VFT. Attempts to fit the data over the entire range by activated dynamics could lead to inconsistent values of the parameters. Finally, we stress that all of this in accord with large dimensionality exact solution of liquid state systems \cite{Maimbourg2016,Kurchan2016,Parisi2020}.

\textbf{\bigskip{}
}

{\bf Acknowledgements:}This work was supported by the National Science
Foundation under Grant Nos PHY 17-08128 and CHE 19-00033, and Welch Foundation (F-0019).

%\bibliography{Glasses1}

\newpage

\begin{thebibliography}{65}
\providecommand{\natexlab}[1]{#1}
\providecommand{\url}[1]{\texttt{#1}}
\expandafter\ifx\csname urlstyle\endcsname\relax
  \providecommand{\doi}[1]{doi: #1}\else
  \providecommand{\doi}{doi: \begingroup \urlstyle{rm}\Url}\fi

\bibitem[Kirkpatrick and
  Thirumalai(1995{\natexlab{a}})]{Kirkpatrick95Transport}
T.R. Kirkpatrick and D.~Thirumalai.
\newblock Are disordered spin glass models relevant for the structural glass
  problem?
\newblock \emph{Transp. Theor. and Stat. Phys.}, 24:\penalty0 927--945,
  1995{\natexlab{a}}.

\bibitem[Cavagna({2009})]{Cavagna09PhysRep}
A~Cavagna.
\newblock {Supercooled liquids for pedestrians}.
\newblock \emph{{Phys. Rep.}}, {476}:\penalty0 51--124, {2009}.

\bibitem[Parisi and Zamponi({2010})]{Parisi10RMP}
G~Parisi and F~Zamponi.
\newblock {Mean-field theory of hard sphere glasses and jamming}.
\newblock \emph{{Rev. Mod. Phys.}}, {82}:\penalty0 789--845, {2010}.

\bibitem[Berthier and Biroli(2011)]{Berthier11RMP}
L~Berthier and G~Biroli.
\newblock Theoretical perspective on the glass transition and amorphous
  materials.
\newblock \emph{Rev. Mod. Phys.}, 83:\penalty0 587--645, 2011.

\bibitem[Biroli and Bouchaud(2012)]{Biroli12}
G.~Biroli and J.P. Bouchaud.
\newblock The random first-order transition theory of glasses: a critical
  assessment.
\newblock In V.~Lubchenko and P.G. Wolynes, editors, \emph{Structural glasses
  and supercooled liquids: theory, experiment and applications}, pages 31--114.
  Johh-Wiley, 2012.

\bibitem[Kirkpatrick and Thirumalai(2015)]{KirkpatrickRMP2015}
T.~R. Kirkpatrick and D.~Thirumalai.
\newblock Random first order concepts in biology and physics.
\newblock \emph{Rev. Mod. Phys.}, 87:\penalty0 183--209, 2015.

\bibitem[Kirkpatrick et~al.(1989)Kirkpatrick, Thirumalai, and
  Wolynes]{Kirkpatrick89PRA}
T.~R. Kirkpatrick, D.~Thirumalai, and P.~G. Wolynes.
\newblock Scaling concepts for the dynamics of viscous liquids near an ideal
  glassy state.
\newblock \emph{Phys. Rev. A}, 40:\penalty0 1045--1054, 1989.

\bibitem[Kirkpatrick and Thirumalai(1987{\natexlab{a}})]{Kirkpatrick87PRL}
T.~R. Kirkpatrick and D.~Thirumalai.
\newblock Dynamics of the structural glass transition and the $p$-spin
  interaction spin-glass model.
\newblock \emph{Phys. Rev. Lett.}, 58:\penalty0 2091--2094, 1987{\natexlab{a}}.

\bibitem[Kirkpatrick and Thirumalai(1987{\natexlab{b}})]{Kirkpatrick87PRB}
T.~R. Kirkpatrick and D.~Thirumalai.
\newblock \textit{p} -spin-interaction spin-glass models: Connections with the
  structural glass problem.
\newblock \emph{Phys. Rev. B}, 36:\penalty0 5388--5397, 1987{\natexlab{b}}.

\bibitem[Kirkpatrick and Wolynes(1987)]{Kirkpatrick87PRBPotts}
T.~R. Kirkpatrick and P.~G. Wolynes.
\newblock Stable and metastable states in mean-field potts and structural
  glasses.
\newblock \emph{Phys. Rev. B.}, 36:\penalty0 8552--8564, 1987.

\bibitem[Kirkpatrick and Thirumalai(1988{\natexlab{a}})]{Kirkpatrick88PRBPotts}
T.~R. Kirkpatrick and D.~Thirumalai.
\newblock Mean-field soft-spin potts glass model: Statics and dynamics.
\newblock \emph{Phys. Rev. B}, 37:\penalty0 5342--5350, 1988{\natexlab{a}}.

\bibitem[Kirkpatrick and
  Thirumalai(1995{\natexlab{b}})]{Kirkpatrick95JPhysiqueI}
T.~R. Kirkpatrick and D.~Thirumalai.
\newblock The cavity approach for metastable glassy states near random
  first-order phase transitions.
\newblock \emph{J. de Physique I}, 5:\penalty0 777--786, 1995{\natexlab{b}}.

\bibitem[Kirkpatrick and Thirumalai(1989)]{Kirkpatrick89JPhysA}
T.~R. Kirkpatrick and D.~Thirumalai.
\newblock Random solutions from a regular density functional hamiltonian: A
  static and dynamical theory of the structural glass transition.
\newblock \emph{J. Phys. A}, 22:\penalty0 L149--L155, 1989.

\bibitem[Mezard and Parisi({1996})]{Mezard96JPhysA}
M.~Mezard and G~Parisi.
\newblock {A tentative replica study of the glass transition}.
\newblock \emph{{J. Phys. A}}, {29}:\penalty0 6515--6524, {1996}.

\bibitem[Mezard and Parisi(2000)]{Mezard00JPhysCondMatt}
M~Mezard and G~Parisi.
\newblock {Statistical physics of structural glasses}.
\newblock \emph{{J. Phys. Cond. Matt.}}, 12:\penalty0 6655--6673, {JUL 24}
  2000.

\bibitem[Bouchaud and Biroli(2004)]{Bouchaud04JCP}
J.P. Bouchaud and G.~Biroli.
\newblock On the adam-gibbs-kirkpatrick-thirumalai-wolynes scenario for the
  viscosity increase in glasses.
\newblock \emph{J. Chem. Phys.}, 121:\penalty0 7347--7354, 2004.

\bibitem[Toninelli et~al.(2005)Toninelli, Wyart, Berthier, Biroli, and
  Bouchaud]{Toninelli05PRE}
C.~Toninelli, M.~Wyart, L.~Berthier, G.~Biroli, and J.-P Bouchaud.
\newblock Dynamical susceptibility of glass formers: Contrasting the
  predictions of theoretical scenarios.
\newblock \emph{Phys. Rev. E}, 71:\penalty0 041505, 2005.

\bibitem[Franz et~al.(2012)Franz, Jacquin, Parisi, Urbani, and
  Zamponi]{Franz12PNAS}
S~Franz, H~Jacquin, G~Parisi, P~Urbani, and F~Zamponi.
\newblock {Quantitative field theory of the glass transition}.
\newblock \emph{Proc. Natl. acad. Sci.}, 109:\penalty0 18725--18730, 2012.

\bibitem[Franz et~al.(2013)Franz, Jacquin, Parisi, Urbani, and
  Zamponi]{Franz13JCP}
Silvio Franz, Hugo Jacquin, Giorgio Parisi, Pierfrancesco Urbani, and Francesco
  Zamponi.
\newblock {Static replica approach to critical correlations in glassy systems}.
\newblock \emph{{J. Chem. Phys.}}, 138:\penalty0 {12A540}, 2013.

\bibitem[Leutheusser({1984})]{Leutheusser84PRA}
E~Leutheusser.
\newblock Dynamical model of the liquid-glass transition.
\newblock \emph{{Phys. Rev. A.}}, 29\penalty0 ({5}):\penalty0 2765--2773,
  {1984}.

\bibitem[Bengtzelius et~al.({1984})Bengtzelius, Goetze, and
  Sjolander]{Bengtzelius84JPhysC}
U~Bengtzelius, W~Goetze, and A~Sjolander.
\newblock Dynamics of supercooled liquids and the glass-transition.
\newblock \emph{{J. Phys. C}}, {17}:\penalty0 5915--5934, {1984}.

\bibitem[Thirumalai and Kirkpatrick({1988})]{Thirumalai88PRB}
D.~Thirumalai and T.~R. Kirkpatrick.
\newblock {Mean-Field Potts Glass Model - Initial condition effects on dynamics
  and properties of metastable states}.
\newblock \emph{{Phys. Rev. B}}, {38}:\penalty0 4881--4892, {1988}.

\bibitem[Das(2004)]{Das04RMP}
S.P. Das.
\newblock {Mode-coupling theory and the glass transition in supercooled
  liquids}.
\newblock \emph{{Rev. Mod. Phys.}}, 76:\penalty0 785--851, 2004.

\bibitem[Kirkpatrick and Thirumalai(1988{\natexlab{b}})]{Kirkpatrick88PRA}
T.~R. Kirkpatrick and D.~Thirumalai.
\newblock Comparison between dynamical theories and metastable states in
  regular and glassy mean-field spin models with underlying first-order-like
  phase transitions.
\newblock \emph{Phys. Rev. A}, 37:\penalty0 4439--4448, 1988{\natexlab{b}}.

\bibitem[Donati et~al.(2002)Donati, Franz, Glotzer, and
  Parisi]{Donati02JNon-Cryst}
C~Donati, S~Franz, S.~C. Glotzer, and G~Parisi.
\newblock Theory of non-linear susceptibility and correlation length in glasses
  and liquids.
\newblock \emph{J. Non-Cryst. Solids}, 307:\penalty0 215--224, 2002.

\bibitem[Thirumalai and Mountain({1993})]{Thirumalai93PRE}
D.~Thirumalai and R.~D. Mountain.
\newblock {Activated dynamics, loss of ergodicity,and transport in supercooled
  liquids}.
\newblock \emph{{Phys. Rev. E}}, {47}:\penalty0 479--489, {1993}.

\bibitem[Flenner and Szamel(2010)]{Flenner10PRL}
E.~Flenner and G.~Szamel.
\newblock Dynamic heterogeneity in a glass forming fluid: Susceptibility,
  structure factor, and correlation length.
\newblock \emph{Phys. Rev. Lett.}, 105:\penalty0 217801, 2010.

\bibitem[Flenner et~al.(2011)Flenner, Zhang, and Szamel]{Flenner11PRE}
E.~Flenner, M.~Zhang, and G.~Szamel.
\newblock Analysis of a growing dynamic length scale in a glass-forming binary
  hard-sphere mixture.
\newblock \emph{Phys. Rev. E}, 83:\penalty0 051501, 2011.

\bibitem[Glotzer et~al.(2000)Glotzer, Novikov, and Schroder]{Glotzer00JCP}
S.~C. Glotzer, V.~N. Novikov, and T.~B. Schroder.
\newblock Time-dependent, four-point density correlation function description
  of dynamical heterogeneity and decoupling in supercooled liquids.
\newblock \emph{The Journal of Chemical Physics}, 112:\penalty0 509--512, 2000.

\bibitem[Maimbourg et~al.(2016)Maimbourg, Kurchan, and Zamponi]{Maimbourg2016}
T.~Maimbourg, J.~Kurchan, and F.~Zamponi.
\newblock Solution of the dynamics of liquids in the large-dimensional limit.
\newblock \emph{Phys. Rev. Lett.}, 116:\penalty0 015902, 2016.

\bibitem[Kurchan et~al.(2016)Kurchan, Maimbourg, and Zamponi]{Kurchan2016}
J.~Kurchan, T.~Maimbourg, and F.~Zamponi.
\newblock Statics and dynamics of infinte-dimensional liquids and glasses: a
  parallel and compact derivation.
\newblock \emph{Journal of Statistical Mechanics: Theory and Experiment},
  2016:\penalty0 033210, 2016.

\bibitem[Parisi et~al.(2020)Parisi, Urbani, and Zamponi]{Parisi2020}
Giorgio Parisi, Pierfrancesco Urbani, and Francesco Zamponi.
\newblock \emph{Theory of simple glasses:exact solutions in infinite
  dimension}.
\newblock Cambridge, 2020.

\bibitem[Szamel(2022)]{Szamel2022}
Grzegorz Szamel.
\newblock An alternative, dynamical density functional-like theory for
  time-dependent density fluctuations in glass-forming liquids.
\newblock \emph{arXiv:2203.09543}, 2022.

\bibitem[Charbonneau et~al.(2017)Charbonneau, Kurchan, Parisi, Urbani, and
  Zamponi]{Charbonneau17AnnRevCondMatPhys}
P.~Charbonneau, J.~Kurchan, G.~Parisi, P.~Urbani, and F.~Zamponi.
\newblock Glasses and jammin transitions: From exact results to
  finite-dimensional descriptions.
\newblock \emph{Ann. Rev. Cond. Mat. Phys.}, 8:\penalty0 265--288, 2017.

\bibitem[Gardner(1985)]{Gardner95NucPhysB}
E.~Gardner.
\newblock Spin glasses with p-spin interactions.
\newblock \emph{Nuc. Phys. B}, 257:\penalty0 747--765, 1985.

\bibitem[Kurchan et~al.(2013)Kurchan, Parisi, Urbani, and
  Zamponi]{Kurchan13JPCB}
Jorge Kurchan, Giorgio Parisi, Pierfrancesco Urbani, and Francesco Zamponi.
\newblock {Exact Theory of Dense Amorphous Hard Spheres in High Dimension. II.
  The High Density Regime and the Gardner Transition}.
\newblock \emph{{J. Phys. Chem. B}}, 117:\penalty0 12979--12994, 2013.

\bibitem[Goldstein(1969)]{Goldstein69JCP}
M~Goldstein.
\newblock Viscous liquids and glass transition - a potential energy barrier
  picture.
\newblock \emph{J. Chem. Phys.}, 51:\penalty0 3728--3729, 1969.

\bibitem[Goldstein(2010)]{Goldstein2010}
M~Goldstein.
\newblock Comparison of activated barriers for the johari-goldstein and alpha
  relaxations and its implications.
\newblock \emph{J. Chem. Phys.}, page 041104, 2010.

\bibitem[Novikov and Sokolov({2003})]{Novikov03PRE}
VN~Novikov and AP~Sokolov.
\newblock {Universality of the dynamic crossover in glass-forming liquids: A
  ``magic{''} relaxation time}.
\newblock \emph{{Phys. Rev. E}}, {67}:\penalty0 031507, {2003}.

\bibitem[Monasson({1995})]{Monasson95PRL}
R~Monasson.
\newblock {Structural glass transition and the entropy of metastable states}.
\newblock \emph{{Phys. Rev. Lett.}}, {75}:\penalty0 2847--2850, {1995}.

\bibitem[Huse and Fisher(1987)]{Huse87JPhysA}
D.A. Huse and D.S. Fisher.
\newblock Pure states in spin glasses.
\newblock \emph{{J. Phys. A.}}, 20\penalty0 (15):\penalty0 L997--L1003, 1987.

\bibitem[Adam and Gibbs(1965)]{Adam65JCP}
G.~Adam and J.H. Gibbs.
\newblock On temperature dependence of cooperative relaxation properties in
  glass-forming liquids.
\newblock \emph{{J. Chem. Phys.}}, 43:\penalty0 139--146, 1965.

\bibitem[Capaccioli et~al.(2008)Capaccioli, Ruocco, and
  Zamponi]{Capaccioli08JPCB}
S~Capaccioli, G~Ruocco, and F~Zamponi.
\newblock {Dynamically correlated regions and configurational entropy in
  supercooled liquids}.
\newblock \emph{{J. Phys. Chem. B}}, 112:\penalty0 10652--10658, 2008.

\bibitem[Fisher and Berker(1982)]{Fisher84PRB}
M.E. Fisher and A.N. Berker.
\newblock {Scaling for 1ST-order phase transition in thermodynamic and finite
  systems}.
\newblock \emph{Phys. Rev. B}, 26\penalty0 ({5}):\penalty0 2507--2513, 1982.

\bibitem[Ferdinand and Fisher(1969)]{Ferdinand69PR}
A.E. Ferdinand and M.E. Fisher.
\newblock {Bounded and inhomogeneous Ising Models. I. Specific heat anomaly of
  a finite lattice}.
\newblock \emph{Phys. Rev.}, 185\penalty0 (2):\penalty0 832--846, 1969.

\bibitem[Imry(1980)]{Imry80PRB}
Y.~Imry.
\newblock Finite-size rounding of a 1st-order phase transition.
\newblock \emph{{Phys. Rev. B}}, {21}:\penalty0 2042--2043, 1980.

\bibitem[Berthier and Kob(2012)]{Berthier12PRE}
Ludovic Berthier and Walter Kob.
\newblock {Static point-to-set correlations in glass-forming liquids}.
\newblock \emph{Phys. Rev. E}, 85:\penalty0 011102, 2012.

\bibitem[Langer(2013)]{Langer13PRE}
J.~S. Langer.
\newblock Ising model of a glass transition.
\newblock \emph{Phys. Rev. E.}, 88:\penalty0 {012122}, 2013.

\bibitem[Villain(1985)]{Villain85JPhysique}
J~Villain.
\newblock Equilibrium critical properties of random field systems - new
  conjectures.
\newblock \emph{J. de. Physique}, 46\penalty0 (11):\penalty0 1843--1852, 1985.

\bibitem[Fisher and Huse(1988)]{Fisher88PRB}
D.S. Fisher and D.A. Huse.
\newblock Non-equilibrium dynamics of spin glasses.
\newblock \emph{{Phys. Rev. B}}, 38:\penalty0 373--385, 1988.

\bibitem[Singh et~al.(1985)Singh, Stoessel, and Wolynes]{Singh85PRL}
Y~Singh, JP~Stoessel, and PG~Wolynes.
\newblock Hard-sphere glass and the density-functional theory of aperiodic
  crystals.
\newblock \emph{{Phys. Rev. Lett.}}, 54:\penalty0 1059--1062, 1985.

\bibitem[Dasgupta and Valls(1999)]{Dasgupta99PRE}
C.~Dasgupta and O.T. Valls.
\newblock {Free energy landscape of a dense hard-sphere system}.
\newblock \emph{{Phys. Rev. E.}}, 59:\penalty0 3123--3134, 1999.

\bibitem[Ramakrishnan and Yussouff(1979)]{Ramakrishnan79PRB}
T.V. Ramakrishnan and M.~Yussouff.
\newblock 1st-principles order-parameter theory of freezing.
\newblock \emph{Phys. Rev. B}, 19:\penalty0 2775--2794, 1979.

\bibitem[Mosayebi et~al.({2010})Mosayebi, Del~Gado, Ilg, and
  Ottinger]{Mosayebi10PRL}
M.~Mosayebi, E.~Del~Gado, P.~Ilg, and H.~C. Ottinger.
\newblock {Probing a Critical Length Scale at the Glass Transition}.
\newblock \emph{{Phys. Rev. Lett.}}, {104}:\penalty0 205704, {2010}.

\bibitem[Tanaka et~al.(2010)Tanaka, Kawasaki, Shintani, and
  Watanabe]{Tanaka10NATMAT}
H.~Tanaka, T.~Kawasaki, H.~Shintani, and K.~Watanabe.
\newblock Critical-like behaviour of glass-forming liquids.
\newblock \emph{Nat. Mater.}, 9:\penalty0 324--331, 2010.

\bibitem[Gibbs and Dimarzio(1958)]{Gibbs58JCP}
J.H. Gibbs and E.A. Dimarzio.
\newblock Nature of the glass transition and the glassy state.
\newblock \emph{{J. Chem. Phys.}}, 28\penalty0 (3):\penalty0 373--383, 1958.

\bibitem[Tanaka(2022)]{Tanaka2022}
Hajime Tanaka.
\newblock Roles of liquid structural ordering in glass transition,
  crystallization, and water's anomalies.
\newblock \emph{J. Non-Cryst. Solids:X}, 13:\penalty0 100076, 2022.

\bibitem[Franz et~al.(2011)Franz, Parisi, Ricci-Tersenghi, and
  Rizzo]{Franz11EPJE}
S.~Franz, G.~Parisi, F.~Ricci-Tersenghi, and T.~Rizzo.
\newblock {Field theory of fluctuations in glasses}.
\newblock \emph{{Eur. Phys. J. E}}, 34:\penalty0 102, 2011.

\bibitem[Thirumalai et~al.({1989})Thirumalai, Mountain, and
  Kirkpatrick]{Thirumalai89PRA}
D~Thirumalai, R.~D Mountain, and T.~R Kirkpatrick.
\newblock {Ergodic Behavior in supercooled liquids and in glasses}.
\newblock \emph{{Phys. Rev. A}}, {39}:\penalty0 3563--3574, {1989}.

\bibitem[Kob and Berthier(2013)]{Kob13PRL}
W.~Kob and L.~Berthier.
\newblock {Probing a Liquid to Glass Transition in Equilibrium}.
\newblock \emph{Phys. Rev. Lett.}, 110:\penalty0 245702, 2013.

\bibitem[Karmakar and Parisi(2013)]{Karmakar13PNAS}
S~Karmakar and G~Parisi.
\newblock {Random pinning glass model}.
\newblock \emph{{Proc. Natl. Acad. Sci.}}, 110:\penalty0 2752--2757, 2013.

\bibitem[Thirumalai(1995)]{Thirumalai95JPI}
D.~Thirumalai.
\newblock {From Minimal Models to Real Proteins: Time Scales for Protein
  Folding Kinetics}.
\newblock \emph{J. Phys. I (Fr.)}, 5:\penalty0 1457--1467, 1995.

\bibitem[Miyagawa et~al.({1988})Miyagawa, Hiwatari, Bernu, and
  Hansen]{Miyagawa88JCP}
H~Miyagawa, Y~Hiwatari, B~Bernu, and J.P Hansen.
\newblock {Molecular-dynamics study of binary soft-sphere mixtures - Jump
  motions of atoms in the glassy state}.
\newblock \emph{{J. Chem. Phys.}}, {88}:\penalty0 3879--3886, {1988}.

\bibitem[Kob and Andersen(1995)]{Kob95PRE}
W.~Kob and H.C. Andersen.
\newblock Testing mode-coupling theory for a supercooled binary lennard-jones
  mixture - the van hove correlation-function.
\newblock \emph{Phys. Rev. E}, 51:\penalty0 4626--4641, 1995.

\bibitem[Kang et~al.(2013)Kang, Kirkpatrick, and Thirumalai]{Kang13PRE}
H.~Kang, T.~R. Kirkpatrick, and D.~Thirumalai.
\newblock {Manifestation of random first-order transition theory in Wigner
  glasses}.
\newblock \emph{Phys. Rev. E.}, 88:\penalty0 042308, 2013.

\end{thebibliography}
%\bibliographystyle{unsrt}

\newpage

%\begin{center}
%\textbf{\large{Figure Captions}}
%\end{center}

%\textbf{Figure 1:} Key predictions of the RFOT theory of the liquid to glass transition. Schematic sketch of the dependence of the $\alpha$ relaxation time on the inverse temperature. Above $T_d$ particle motion occurs by a diffusive process. Activated transition, marked by $\tau_{\alpha} \approx 10^{-7}$ (\cite{Novikov03PRE}), occurs below $T_d$. The temperature dependence of $\tau_{\alpha}$ till $T_d$ follows a power law. Below $T_d$, $\tau_{\alpha}$ increases faster than predicted by Arrhenius law, and can be captured using the VFT law.  The inverse of the laboratory glass transition temperature is defined by the condition that shear viscosity is $\approx 10^{13}$ poise. At the inverse equilibrium transition temperature, $T_K$, the relaxation time diverges. In the temperature range $T_K \le T \le T_d$ the correlation length, depicted by the compact droplets,increase eventually diverging at $T_K$ with an exponent $2/d$. Relaxation times for several glass forming materials can be analyzed using the RFOT predictions.  

%\newpage	
%\begin {figure}[t]
%\centering 
%\includegraphics[width=1.0\columnwidth]{schematic.pdf}  
%\end{figure}   

\end{document}